\documentclass{article}

\usepackage{arxiv}

\usepackage[utf8]{inputenc} 
\usepackage[T1]{fontenc}    
\usepackage{hyperref}       
\usepackage{url}            
\usepackage{booktabs}       
\usepackage{amsfonts}       
\usepackage{nicefrac}       
\usepackage{microtype}      
\usepackage{lipsum}
\usepackage{graphicx}
\usepackage{amsmath}
\usepackage{amssymb}
\usepackage[table,xcdraw]{xcolor}
\usepackage{multirow}
\usepackage{framed,multirow}

\usepackage{amssymb}
\usepackage{latexsym}
\usepackage{amsmath}
\usepackage{url}
\usepackage{xcolor}
\usepackage{float}
\usepackage{color}
\usepackage[normalem]{ulem}

\usepackage{hyperref}

\title{DeepLesionBrain: Towards a broader deep-learning generalization for multiple sclerosis lesion segmentation}

\author{
  Reda Abdellah Kamraoui\\
   Univ. Bordeaux, Bordeaux INP,\\
  CNRS, LaBRI, UMR5800, PICTURA,\\
  F-33400 Talence, France\\
  \And
 Vinh-Thong Ta\\
  Univ. Bordeaux, Bordeaux INP,\\
  CNRS, LaBRI, UMR5800, PICTURA,\\
  F-33400 Talence, France\\

  \And
 Thomas Tourdias\\
 Service de Neuroimagerie Diagnostique et Thérapeutique,\\
 Univ. Bordeaux, INSERM, Neurocentre Magendie,\\
 U1215, F-3300 Bordeaux, France\\

\And
 Boris Mansencal\\
  Univ. Bordeaux, Bordeaux INP,\\
  CNRS, LaBRI, UMR5800, PICTURA,\\
  F-33400 Talence, France\\
  
  \And
 José V Manjon\\
  ITACA, Universitat Politècnica de València,\\
  46022 Valencia, Spain\\

  \And
   Pierrick Coupé\\
   Univ. Bordeaux, Bordeaux INP,\\
  CNRS, LaBRI, UMR5800, PICTURA,\\
  F-33400 Talence, France\\
}

\begin{document}
\maketitle

\begin{abstract}
Recently, segmentation methods based on Convolutional Neural Networks (CNNs) showed promising performance in automatic Multiple Sclerosis (MS) lesions segmentation. These techniques have even outperformed human experts in controlled evaluation conditions such as Longitudinal MS Lesion Segmentation Challenge (ISBI Challenge). However state-of-the-art approaches trained to perform well on highly-controlled datasets fail to generalize on clinical data from unseen datasets. 
Instead of proposing another improvement of the segmentation accuracy, we propose a novel method robust to domain shift and performing well on unseen datasets, called DeepLesionBrain (DLB). This generalization property results from three main contributions. First, DLB is based on a large group of compact 3D CNNs. This spatially distributed strategy ensures a robust prediction despite the risk of generalization failure of some individual networks. Second, DLB includes a new image quality data augmentation to reduce dependency to training data specificity (\textit{e.g.}, acquisition protocol). Finally, to learn a more generalizable representation of MS lesions, we propose a hierarchical specialization learning (HSL). HSL is performed by pre-training a generic network over the whole brain, before using its weights as initialization to locally specialized networks. 
By this end, DLB learns both generic features extracted at global image level and specific features extracted at local image level. DLB generalization was validated in cross-dataset experiments on MSSEG'16, ISBI challenge, and in-house datasets. During experiments, DLB showed higher segmentation accuracy, better segmentation consistency and greater generalization performance compared to state-of-the-art methods. Therefore, DLB offers a robust framework well-suited for clinical practice.
\end{abstract}


\section{Introduction}
In recent years, the medical imaging community has witnessed a rapid increase in image processing methods based on Deep Learning (DL). These novel techniques came with remarkable performance in many tasks including Multiple Sclerosis (MS) lesion segmentation. Some automated algorithms have even reached human level performance in controlled evaluation conditions (see \cite{carass2017longitudinal}).
Unlike over-controlled conditions where most DL approaches have been validated, real world data exhibit high diversity. Consequently, clinical use of MS lesion segmentation based on DL is still limited mainly because of their poor generalization on new data coming from medical sites that have not been covered during training (unseen domains). This lack of generalization of DL methods can result from several factors such as the selected solution during optimization, the diversity of the training dataset or the genericity of the extracted features.

DL is based on the assumption that training and test data are independent but come from the same distribution. 
This assumption is usually not respected in medical imaging data especially for Magnetic Resonance Imaging (MRI) where acquisition protocols, MRI scanner, patient populations, and software processing may vary depending on the clinical center or the cohort. As a result of these differences of data distribution (covariate shift), a decrease in performance is observed between the training data (source domain) and the test data derived from different distributions (target domain). This is known as the domain shift. 

An intuitive method to reduce this problem is to train on a wider and more heterogeneous dataset (as shown by \cite{maartensson2020reliability}). However this requires a large dataset annotated by experts which is rarely available and tedious to produce.
Some deal with this phenomenon by applying extensive data augmentation (such as \cite{zhang2020generalizing}). Others use few available labeled images from the target domain in order to reduce the covariate shift, such as few-shot and one-shot learning strategies (see \cite{snell2017prototypical,valverde2019one}).

Besides, DL requires the tuning of a large number of parameters relatively to the number of training data samples. Thus, it usually ends up converging to one of the many possible local minima as opposed to the theoretical best parameter configuration which leads to the global minimum.  
Consequently, the generalization ability of the model depends on the selected solution. The selection of the best generalizing local minimum is still an open question. On one hand, some works have proposed to select it using the local characteristics (\textit{e.g.}, flatness) of the loss function (see \cite{keskar2016large, wu2017towards} ). On the other hand, an alternative strategy consists in combining several local minima to improve the generalization capability of the method. This can be done by averaging several local minima of one model (\textit{e.g.}, snapshot ensemble \cite{huang2017snapshot}) or by combining outputs of different models trained independently (\textit{e.g.}, classical ensemble \cite{zhang2019multiple} and spatially distributed networks \cite{coupe2020assemblynet, huo20193d}).

Unlike classical methods that use hand crafted features, Convolutional Neural Networks (CNNs) automatically extract the most suitable set of features automatically for a particular task. Although this strategy is very efficient to extract relevant features for a particular source domain, this set of features may not generalize well for the target domain. Some works proposed to learn invariant features that coexist across different source domains \cite{motiian2017unified,muandet2013domain,yang2013multi}. They tried to apply a regularization in order to learn an abstract representation of the specific computer vision task (\textit{i.e.}, just like humans understand high level concepts). Indeed, the extraction of generalizing features lies between the freedom of the optimization process to find the optimal combination from data, and the constraints used for minimizing domain bias.

The successful deployment of DL based methods for MS lesion segmentation requires generalization capabilities that can guarantee high performance for unseen domains. First, such methods should ensure the convergence of the DL model to generalizing minima. Second, the training process should anticipate the reduction of the covariate shift.  Moreover, the method should be enforced to learn MS lesion generalizing features from the source domain, in order to effectively delineate lesions despite the target domain distribution. Finally, this solution should not require additional annotation in case of processing unseen domains.

Recently, many works have been proposed for MS lesion segmentation using CNNs.\\ 
First, \cite{brosch2016deep} proposed a deep 3D encoder-decoder network, with joint training of the encoder and the decoder. The authors used shortcut connections between the two interconnected pathways for integrating high and low level features. This pioneer work demonstrated the high potential of deep learning for MS lesion segmentation.\\
\cite{valverde2017improving} proposed a cascade of two patch-wise 3D CNNs, composed of a first sensitive network to reveal possible lesion candidates followed by a second network to reduce misclassified voxels. This cascade allows refined segmentation but it uses a small receptive field that prevents capturing the global context. Later, the authors \cite{valverde2019one} improved their method by proposing an one-shot domain adaptation model which uses transfer learning and partial fine-tuning. However, this domain adaptation needs a labeled example from the new domain. Moreover, such strategies lead to different versions of the method after each adaptation, this results in discrepancies in the segmentation. \\
\cite{hashemi2018asymmetric} considered the problem of data imbalance (\textit{i.e.}, the under-sampling of the lesion class) by using an asymmetric similarity loss function based on Tversky index to train a 3D CNN that performed better than Dice or cross entropy measures. This result suggests that further work should be done on choosing adequate loss function. Although the proposed loss can be tuned for the optimal trade-off between precision and recall in a particular domain, the generalization to unseen domains have to be proven.\\ 
\cite{zhang2019multiple} used a fully convolutional densely connected network for MS lesion segmentation. They stacked adjacent 2D slices of different modalities with a channel-wise concatenation, before forwarding this stack through a 2D CNN. The final segmentation is based on a majority vote along different orientations. While this method showed competitive performance on a well-controlled challenge, the stacking using only the two directly adjacent slices gives a weak insight on the 3D nature of the data. 
Moreover, 2D features may not be considered as generalizing features when processing 3D volumes and can result in the limited generalization of the method.\\
\cite{aslani2019multi} proposed an end-to-end encoder-decoder 2D network with multiple downsampling branches, one for each input modality, and a decoder part where features from the different modalities are put together at multiple scales. This separation in encoder branches enables the model to encode information efficiently from each modality, before combining them in a later stage. However this 2D approach does not combine features based on axial, coronal, and sagittal orientations that may greatly reduce its generalization on 3D images.\\
\cite{feng2019self} considered MRI modality unavailability during segmentation by introducing sequence dropout. 
This is an important point since the availability of all the modalities is not always ensured between datasets that can greatly reduce the generalization capacity of a method. This framework randomly drops specific MRI sequences during training, with the intent to learn the intrinsic information of each sequence. This technique showed it can produce acceptable segmentation even in the absence of one or two modalities. Nonetheless, it is less efficient than other state-of-the-art methods when all modalities are available (will be detailed in section \ref{same domain val}).\\
\cite{aslani2020scanner} tackled the problem of generalization to new domains by integrating a regularization network to the traditional encoder-decoder network. The regularizer penalizes the network when the later learns features which allow the prediction of MRI scanning sites. However, \cite{li2018domain} have argued that such strategies suffer from overfitting, the obtained representation could well generalize for all the source domains but poorly for the unknown target domains.

All the cited MS methods (\cite{brosch2016deep,valverde2017improving,feng2019self,hashemi2018asymmetric,zhang2019multiple,aslani2019multi,feng2019self}) focused on obtaining accurate segmentation within a same domain evaluation. 
However, the use of out-of-domain datasets is essential to ensure a good evaluation of the generalization capabilities of a method. This question is right now a hot topic (see \cite{maartensson2020reliability}, and \cite{bron2020cross}) and an important recommendation from the clinical world (see \cite{omoumi2021buy}). Experiments using training and testing images derived from the same domain are known to be biased [\cite{omoumi2021buy}] and do not ensure generalization.
Therefore, a model used in clinical conditions should produce accurate segmentation for new domain images without the need of retraining with expert segmentation on the new domain.

In this paper we propose DeepLesionBrain (DLB), a novel method for MS lesion segmentation robust to domain shift, validated on out-of-domain testing (cross-dataset testing).\\ 
First, we use a large group of compact 3D CNNs spatially distributed over the brain with overlapping receptive fields between regions. 
By associating a distinct network with each region of the brain, the spatially distributed networks (see \cite{coupe2020assemblynet, huo20193d}) 
strategy simplifies the MS lesion segmentation from a single complex task on the whole brain to multiple simpler sub-tasks on each region.  
Moreover, the overlapping regions ensure consistent and stable consensus.
Studies highlighted that such consensus-based strategy helps the convergence towards generalizing minima (see  \cite{breiman2001random,izmailov2018averaging,olson2018modern}).
Moreover, applied to brain segmentation, this strategy demonstrated good generalization on unseen domains (\textit{e.g.}, child's brain or patients with Alzheimer's disease) when trained on healthy adult brains [\cite{coupe2020assemblynet}].
\\
Second, DLB is trained with a novel Image Quality Data Augmentation (IQDA) method, which mimics real world data diversity by adding realistic alterations to the training images. As shown in the works of \cite{zhang2020does}, such a type of regularization technique aims to reduce covariate shift. 
IQDA proposes specific augmentations that constrain task learning to be independent from source data acquisition resolution, data contrast, or data quality. Consequently, the proposed augmentation strategy enables domain shift robustness.
\\
Third, to ensure good features generalization, we focus on feature learning strategy. We consider that a generalizing model should learn two types of features: first global and generic features, and second local and specific features. Therefore, we propose Hierarchical Specialization Learning (HSL) to efficiently extract those features in two steps. In the first step, a single network (the generic network) is trained on all brain regions. In the second step, each network of the spatially distributed networks is initialised with the generic network weights and specialized for a specific region of the brain.
\\
Finally, we propose a method using only two modalities (T1w and FLAIR) to ensure its compatibility with a large number of datasets. 
Most of the methods (\cite{feng2019self,brosch2016deep,valverde2017improving,zhang2019multiple}) optimize their segmentation using T1w, FLAIR, PD, and T2 modalities. However in clinical conditions, not all these sequences are always available. 
Therefore we focused our work on developing a robust approach using only two modalities.

\section{Method and Material}

\begin{figure*}[htbp]
    \centering
    \newcommand{\sz}{1\textwidth}
    \includegraphics[width=\sz]{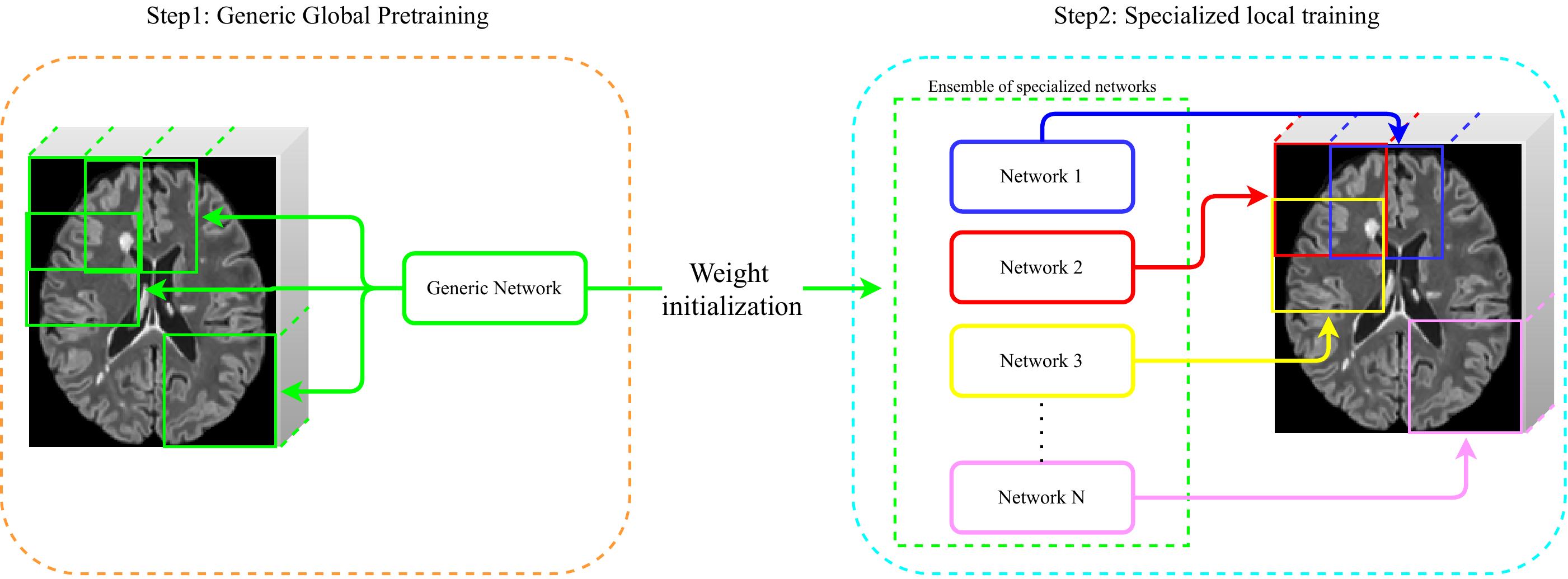}
      \caption{The two steps training process of DeepLesionBrain (see
      Sect.~\ref{HSL} for more details)}
    \label{fig:DLB}
\end{figure*}

\subsection{Method Overview}

\subsubsection{Spatially Distributed Networks Strategy}
DeepLesionBrain (DLB) uses a group of compact networks, each network is specialized in a particular region of the brain, and processes a sub-volume of the global volume. The receptive fields of the neighboring networks overlap with one another, and the final segmentation of the whole volume is obtained with a majority vote of the local predictions. Employing our spatially distributed compact networks is equivalent to a big network with more filters and a higher receptive field (see Fig. \ref{fig:DLB}).
This particular configuration with overlapping receptive fields also permits domain generalization and robustness to failure of an individual network. Indeed, employing a group of spatially distributed networks collectively ensures a stable segmentation. The majority vote on overlapped regions produces a consensus more robust to domain shift even if the prediction of some networks taken separately may be affected by covariate shift.

\subsubsection{Image Quality Data Augmentation (IQDA)} \label{IQDA}

The quality of the MRI greatly varies between datasets. In fact, the quality of the images depends on several factors such as signal to noise ratio, contrast to noise ratio, resolution or slice thickness. \\
To address this issue, we propose a data augmentation strategy which considers image quality disparity. 
During training, we simulate “on the fly” altered versions of 3D patches. We randomly introduce at each iteration either blur, edge enhancement, or axial subsampling distortion (2D FLAIR are usually acquired along the axial direction).
For the blur, a gaussian kernel is used with a randomly selected standard deviation ranging between $[0.5, 1.75]$. For edge enhancement, we use unsharp masking with the inverse of the blur filter. For axial subsampling distortion, we simulate acquisition artifacts that can result from the varying slice thickness. We use a uniform filter (a.k.a mean filter) on the axial direction with a size of $[1 \times 1 \times sz]$ where $sz \in {2,3,4}$. Ground truth is kept the same as the original version.  This process reduces the domain bias when learning to extract relevant features caused by data variability.

\subsubsection{Hierarchical Specialization Learning (HSL)} \label{HSL}
In order to achieve generalization for MS lesion segmentation, we aim to extract generalizing lesion features. These features can be grouped into two categories.
First, some lesion characteristics are considered generic and shared among lesion types.
Such features are independent from lesion localization. They have a common and inherent significance at the global scale of the brain volume, we will refer to them in this paper as “generic global features”.
Second, other relevant features for MS lesions depend on brain structure and some distinct regions (see \cite{filippi2019assessment}). 
In this work we refer to these features as “specialized local features”.

On the one hand, training each specialized network on a specific sub-region of the brain (see Fig. \ref{fig:DLB} right) would prevent the efficient learning of “generic global features”, since each specialized member of our group would be trained on a particular region of the whole brain. 
On the other hand, using a single 3D CNN to learn "specialized local features" over the whole brain volume would require a large model which may not fit into memory and a large dataset to train it.

To overcome this limitation and in order to learn generalizing features, we propose a novel Hierarchical Specialization Learning (HSL).
This two step learning process shown in Fig. \ref{fig:DLB} is equivalent to a cascade of networks with the first layer of the cascade focusing on "generic global features", and the second one on "specialized local features".
First, the “generic network” is trained with data samples from all over brain regions to learn general knowledge about lesions by extracting “generic global features”. 
Second, each network in the spatially distributed strategy is specialized over a specific sub-volume of the brain.

The generic network is used as an initialization for each network of our spatially distributed networks, by transferring the generic network weights to each individual specialized network. 
The knowledge gained from this transfer learning transmits the ability to extract “generic global features”, while the specialized network training will specialize them in extracting local “specialized local features”.

In our ablation study, we will show that this hierarchical specialization learning of the specialized networks performs better than training a single network over the whole brain, or training the specialized networks without HSL. 

\subsubsection{Selection of the Required Modalities}
In order to use a trained model for MS lesion segmentation with optimal performance, this usually requires to use the same set of modalities that have been chosen during training. DLB proposes a method that needs only T1w and FLAIR sequences in order to be compatible with all benchmark MS datasets and most already available MS patients data.

Our method is built with the purpose to generalize on unseen datasets, thus it uses the minimum necessary modalities. Indeed, increasing the number of sequences requires longer scan acquisition time. Besides, it needs more complex processing which may be prone to error, such as multimodal image registration. Furthermore, the use of more sequences during the training on a dataset may reduce the generalization to other image domains.

In addition to the wide use of T1w and FLAIR for MS diagnosis, the choice of these modalities have also been motivated by the fact that FLAIR is the most relevant sequence for revealing most of MS lesions (see \cite{narayana2020multi}), while T1w can provide complementary structural information needed for accurate segmentation.


\begin{figure}[htb!]
  \centering
  \newcommand{\sz}{250pt}
  \includegraphics[width=\sz]{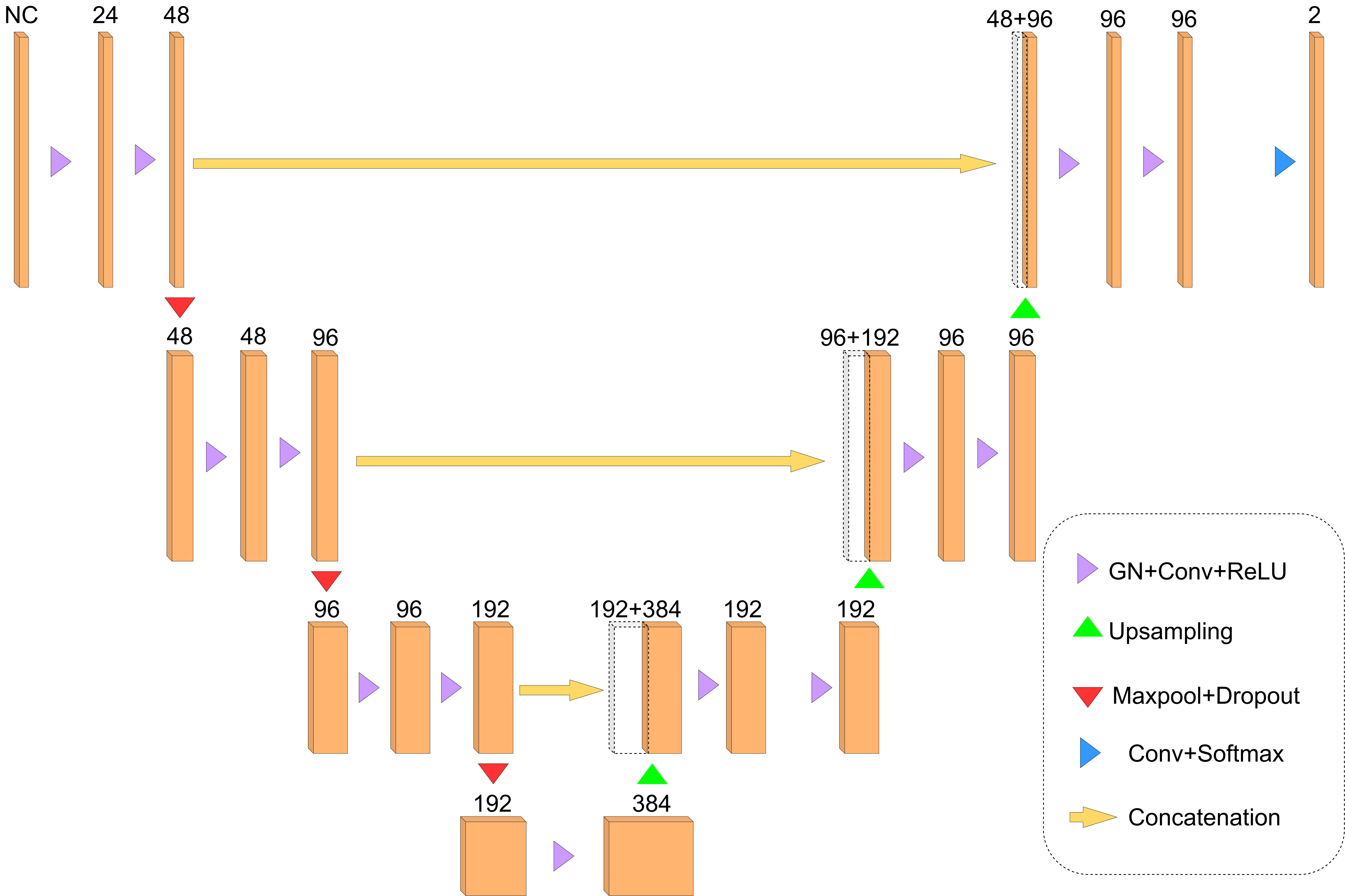}
  \caption{Illustration of the considered U-Net architecture. The number of input channels (NC) depends on the modality number (\textit{i.e.}, NC= 2, for using T1w and FLAIR ). Each block is composed of group normalization (GN), Convolution (Conv) and Rectified Linear Unit (ReLU) activation. }
  \label{fig:dlb-archi}
\end{figure}

\subsection{Implementation Details}
The network architecture used in our spatially distributed strategy is based on 3D U-Net composed of a downsampling part and an upsampling one, linked with one another by skip connections at the multiple scales. This 3D CNN architecture, shown in Figure~\ref{fig:dlb-archi}, has been used for all the networks in our approach.
Dropout with 0.5 rate is used after max-pooling layers to prevent the overfitting of our model to the training data.
Due to GPU memory constraints, we trained with a batch size of 1, and so we used Group Normalization (GN) \cite{wu2018group} with 8 groups before each convolution.
We have chosen Rectified Linear Units (ReLu) to introduce non linearity after convolution layers. 

DLB is optimized with Adam \cite{kingma2014adam} using a learning rate of 0.0001 and a momentum of 0.9.
The experiments have been performed with Keras 2.2.4 \cite{chollet2015keras} and Tensorflow 1.12.0 \cite{abadi2016tensorflow} on NVIDIA Titan Xp 12 GB GPU.

\subsection{Method Description}
To obtain sub-volumes for each image, we first divide our whole MRI registered into the MNI space into multiple overlapping 3D patches. We perform a cropping operation over the whole image using a sliding window of the size $(Px,Py,Pz)$, and a stride of $(Sx,Sy,Sz)$. We take $Sx<Px$,  $Sy<Py$,  $Sz<Pz$ to ensure the overlapping. 

Sub-volumes from different images, that represent the same receptive field into the MNI space (the same sub-volume region of the whole volume), are grouped together. They are used for training a network specialized for that particular region.

In this work, we explored many combinations of sub-volume sizes and numbers. We chose a configuration with 125 sub-volumes by taking experimentally $Px=Py=Pz=96$, $Sx=Sy=76$, and $Sz=67$ as a good trade-off between the overall performance and computation resources.

\subsubsection{Symmetrical Training}\label{Symmetrical Training}
To limit redundant training and to use the most possible data for a particular brain region, we choose to train specialized networks only on one hemisphere. By flipping (mirroring) the sub-volumes of the second hemisphere, it is possible to train a single specialized network for both sides. Thus, we can use twice the amount of data for each region while reducing the number of networks to train to nearly a half (due to sub-volume overlapping, the plane of symmetry cuts through the median networks which cover equivalent symmetrical regions from both hemispheres). Consequently, unlike previous works with spatially distributed networks (\textit{i.e.}, \cite{coupe2020assemblynet}), instead of using 125 networks we use only 75 specialized networks.  We experimentally verified that using 125 networks or only 75 trained with twice the number of patches, produced similar segmentation accuracy.

\subsubsection{Loss Function}
MS lesion segmentation task suffers from class imbalance since lesion volume is considerably low compared to healthy volume. Thus, we use a smooth version of the Generalized Jaccard Loss (GJL), which considers this issue \cite{manjon2020deephips}. 

\begin{equation}
G J L =1-\frac{ \sigma +\sum_{c=1}^{N c} w_{c} \sum_{i=1}^{N} p_{c i} t_{c i} }{\sigma + \sum_{c=1}^{N c} w_{c} \left(\sum_{i=1}^{N} (p_{c i}+t_{c i}) -\sum_{i=1}^{N} p_{c i} t_{c i}\right)}
\end{equation}

Where $w_{c}=1 / (1+\sum_{i=1}^{N} t_{c i})$, $\sigma$ is the smoothness factor, $N$ is the number of voxels, $Nc$ is the number of classes, $p_{c i}$ and $t_{c i}$ are respectively the predicted probability and the ground truth probability of the vowel $i$ for the class $c$.

During inference, we combine the overlapping predictions in a straightforward aggregation majority vote technique. The class of each voxel (either lesion or healthy tissue) is chosen based on the most predicted class among the networks which cover that voxel.

\subsection{Datasets}
To assess the robustness of a model, it is crucial to test its ability to generalize on unseen domains. Therefore, DLB has been trained and validated using different datasets to assess its domain generalization ability (see \ref{Cross Dataset Testing} ). These datasets exhibit high heterogeneity in terms of resolution, data processing, acquisition sites, delineation protocols, and they also cover a large variety of clinical scenarios.  

\subsubsection{ISBI Longitudinal Multiple Sclerosis Lesion}

The ISBI dataset \cite{carass2017longitudinal} consists of five subjects for training, fourteen subjects for testing, with a mean of 4.4 time-points per subject (21 images for training and 61 images for testing). Two human expert raters delineated MS lesions, from the four available modalities acquired on 3.0 Tesla MRI scanner:
3D MPRAGE $T_{1}-$weighted (T1w) of $0.82\times0.82\times1.17$ $mm^{3}$ voxel size,  2D $T_{2}-$weighted (T2), 2D $T_{2}-$weighted fluid attenuated inversion recovery (FLAIR),  and 2D Proton Density weighted (PD), of $0.82\times0.82\times 2.2$ $mm^{3}$ voxel size each.

For the training, we used the ISBI training dataset with available annotations from the two experts. For test and evaluation, we segmented the test data with no available expert annotation, and submitted our results to the ISBI challenge website\footnote{https://smart-stats-tools.org/lesion-challenge-upload-results}.
The ISBI pipeline already included preprossessing. Each first time-point T1w was inhomogeneity-corrected using N4 [\cite{tustison2010n4itk}], skull-stripped \cite{carass2007joint}, dura stripped \cite{shiee2014reconstruction}, followed by a second N4 inhomogeneity correction, and rigid registration to a 1 $mm^{3}$ isotropic MNI template. Then, this image was used as a target for the remaining T1w time-points and all modalities for the same subject. These images were N4 corrected and then rigidly registered to the T1w in the MNI space. The skull and dura-stripped mask from the target T1w was applied, which were then N4 corrected again. We added an intensity normalization step using kernel density estimation for all images.

\subsubsection{MICCAI2016 MS Challenge Dataset}

The MSSEG'16 training dataset \cite{commowick2016msseg} contains 15 patients from 3 different clinical sites. Five modalities are available for each patient: 3D FLAIR, 3D T1w, 3D T1w GADO, 2D PD, and 2D T2.
The images were acquired on 1.5T and 3T MRI scanners with multiple resolutions: 

3D FLAIR modalities ranging from $1\times0.5\times0.5$ to $1.25\times1.04\times1.04$ $mm^{3}$, and 3D T1w sequences between $0.85\times0.74\times0.74$ and $1.08\times1.08\times0.9$ $mm^{3}$. 

Seven human experts have manually segmented the multiple sclerosis lesions.
Each patient modalities have been preprocessed with the same pipeline. First, each sequence was denoised using the non local means algorithm \cite{coupe2008optimized}. Second, a rigid registration of each modality on the FLAIR was performed \cite{commowick2012block}. Then, skull stripping of T1w was performed using the volBrain platform \cite{manjon2016volbrain}, the same mask is applied to other modalities. Finally, bias field correction was applied using the N4 algorithm \cite{tustison2010n4itk}. In addition to these steps that have been performed on the available images, each modality was registered to the MNI space for our experiments.
Similarly to the ISBI images, we used kernel density estimation for the normalization step.

\subsubsection{In-house Dataset}

For further evaluation of our approach, we used an In-house 3D MRI dataset, with 3D T1w and 3D FLAIR modalities \cite{coupe2018lesionbrain}. This dataset contains 43 subjects diagnosed with MS. The images were acquired with different scanners and multiple resolutions ($0.6\times0.6\times0.65$  $mm^{3}$, $0.5\times0.5\times0.9$  $mm^{3}$, and  $1\times1\times1$  $mm^{3}$).

The dataset lesion masks have been obtained by human experts manual delineation. The images were pre-processed using the lesionBrain pipeline from the volBrain platform [\cite{manjon2016volbrain}]. First, it included denoising of each modality [\cite{coupe2008optimized}]. Second, an affine registration to MNI space was performed on the T1w , then the FLAIR was registered to the transformed T1w. Skull stripping and bias correction have been performed on the modalities, followed by a second denoising. Finally, the intensities have been normalized.

\begin{table*}[!ht]\label{tab:datasets}
    \centering
    \caption{Description of datasets used in this work.}
    \label{tab:my-table}
    \begin{tabular}{lccccc}
    \hline
    &
    FLAIR resolution & 
    Site & 
    \# Subjects & 
    \# Raters &
    Modalities\\ 
    \hline
    ISBI train-set   & 2D& Mono& 5, multiple time-points& 2& T1w, FLAIR, T2, PD\\
    MSSEG'16& 3D& Multi& 15& 7& T1w, T1w GADO, FLAIR, PD, T2\\
    In-house& 3D& Multi& 43& 1& T1w, FLAIR\\
    \hline\\
    \end{tabular}
\end{table*}

\subsubsection{Datasets Summary}

Table~\ref{tab:my-table} summarizes the main differences between the 3 datasets. We focused specifically on the resolution of FLAIR due to its known relevance in MS lesion segmentation.
To summarize, ISBI train-set contains multiple time points of only five subjects, acquired in a single clinical site with two human expert segmentations. Except 3D MPRAGE T1w, the other three modalities are in 2D. MSSEG'16 dataset is a multi-site database comprising 15 patients, with seven human segmentations. This dataset contains 5 available modalities with 3D FLAIR. Finally, In-house dataset is the largest dataset with 43 patients, and multi-site 3D modalities, segmented by a single human rater and validated by a second one.

\begin{figure*}[!t]
    \centering
    \newcommand{\sz}{\textwidth}
    \includegraphics[width=\sz]{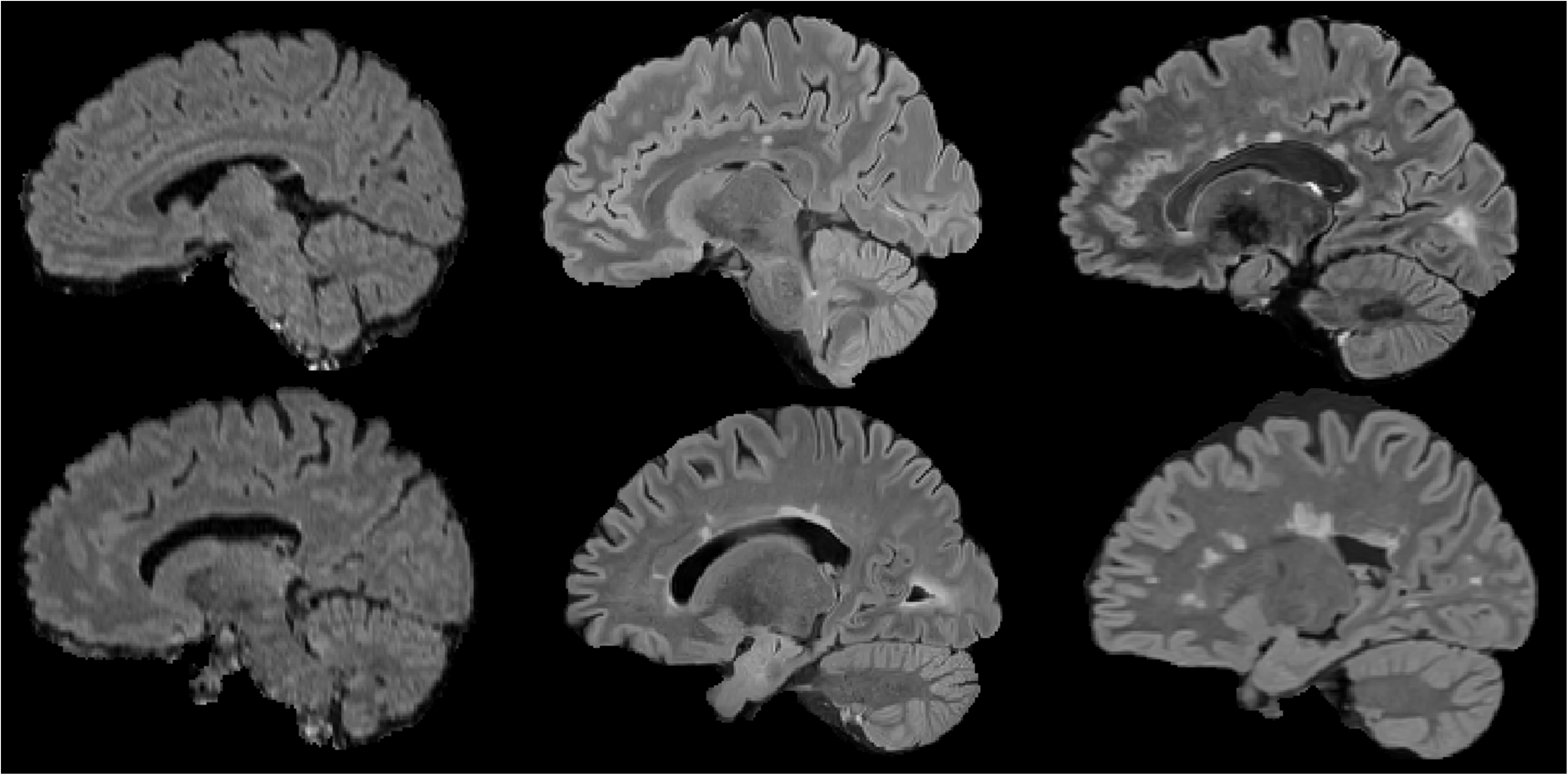}
    \caption{FLAIR examples from the considered three datasets in the MNI space and after intensity normalization. From left to right, the two images are from ISBI, MSSEG'16, our in-house data, respectively.}
    \label{fig:dataset-examples}
\end{figure*}

Figure~\ref{fig:dataset-examples} shows examples from the three presented datasets, each image represents a sagittal section of the FLAIR modality in the MNI space after intensity normalization. The two images on the left are examples from the ISBI dataset. We notice blurring effects which makes it hard to distinguish precisely brain structures. This blur comes from 2D low resolution acquisitions. 
In the middle, the two examples come from the MSSEG'16 dataset. These 3D FLAIR are noticeably of higher resolution than the other images. Therefore, lesion boundaries are more easily delineated and main structures are clearly apparent. The final two images on the right are from our In-house 3D dataset. 
The 3D resolution enables the differentiation of white matter, gray matter, and shows the lesions clearly.

In terms of FLAIR images, we notice that both MSSEG'16 and In-house dataset (3D FLAIR) propose better visual quality than ISBI dataset (2D FLAIR).   

\subsection{Validation Framework}
\subsubsection{Evaluation Metrics}
The assessment of a segmentation method is usually measured by a similarity metric between the predicted segmentation and the human expert ground truth. 

First, we use several complementary metrics in order to assess segmentation performance.
Namely, we use the Dice similarity coefficient, the Positive Predictive Value (PPV or the precision), true positive rate (TPR, known as recall or Sensitivity), and Pearson's correlation coefficient (CORR).

\begin{equation}
P P V=\frac{T P}{T P+F P}
\, \mathrm{,}
\,\,\,\,\,\,\,\,\,\,\,\,\,\,\,\,\,\,\,\,\,\,\,\,
T P R=\frac{T P}{T P+F N}
\, \mathrm{,}
\end{equation}

\begin{equation}
Dice=\frac{2 \times T P}{ (T P+F N) +(T P+F P) }\, \mathrm{,}
\end{equation}

where TP, FN, FP represent respectively true positives, false negatives, and false positives.

Second, recent works (\textit{i.e.}, \cite{commowick2018objective}) question the relevance of classic metrics (Dice) compared to detection metrics, which are used for MS diagnostic and clinical evaluation of the patient evolution. Thus, in addition to the voxel-wise metrics, we also use lesion-wise metrics that focus on the lesion count. Such as, Lesion False Positive Rate (LFPR) and Lesion True Positive Rate (LTPR). 

\begin{equation}
L T P R=\frac{L T P}{L T P+L F N}
\, \mathrm{,} \quad \quad
L F P R=\frac{L F P}{L T P+L T N} \, \mathrm{,}
\end{equation}

where LTP, LFN, LFP represent respectively lesion true positives, lesion false negatives, and lesion false positives.

ground truth volumes.

Moreover, \cite{garcia2013review} pointed out that even though Dice is commonly used and simplifies method comparison, multiple complementary metrics are needed to provide a better understanding of the performance. Recently, international challenges took into consideration several metrics (\cite{carass2017longitudinal} and \cite{commowick2018objective}).  Consequently, we decided to evaluate our methods using Hybrid score proposed by \cite{carass2017longitudinal}. This metric combines voxel-wise segmentation, lesion-wise detection, and volumetric metrics. It is defined as:

\begin{equation}
Hybrid=\frac{Dice}{8}+\frac{P P V}{8}+\frac{(1-L F P R)}{4}+\frac{L T P R}{4}+\frac{C O R R}{4}
\end{equation}

Finally, we also use the ISBI Submission Score for the evaluation of ISBI test-set segmentations. \cite{carass2017longitudinal} defined it as the average of the hybrid scores of all image examples with the different human raters and with inter rater variability taken into consideration. This score is computed after submitting the segmentation to ISBI’s challenge website \footnote{https://smart-stats-tools.org/lesion-challenge}. Obtaining an ISBI score of 90 or higher with a segmentation technique indicates that this method is similar to the human raters.

\subsubsection{Cross-dataset Experiments}
In our context, the domain is characterised by acquisition hardware, data processing, and annotation protocol. Thus, source and target domains are related but slightly different, restricting the applicability of traditional supervised learning models 
which rely on the assumption that train and inference data come from the same distribution .

Both domain generalization and adaptation are concerned with reducing dataset bias.
The difference between these strategies is that for domain adaptation, some unlabeled data or even a few labeled data from the target domain are exploited to capture properties of the target domain for model adaptation. However, in domain generalization no samples of any kind are used from the target domain.

Domain generalization has been proposed to address the problem of unavailability of target domain samples by leveraging the labeled data to learn a universal representation in order to generalize for any target domain and without any prior insight from that domain.

In this work, we emphasize on testing the domain generalization of our approach with cross-dataset evaluation. 
Unlike domain adaptation such as one-shot domain adaptation [Valverde et al., 2019], DLB does not need expert segmentation from the target domain.
Our testing conditions draw a clear distinction between training data containing source domains and testing data containing unseen target domain.

\subsubsection{Reference Methods}

During experiments, our method was compared to 
three publicly available state-of-the-art approaches. We performed training and validation for all three compared methods, in the same conditions regarding datasets and preprocessing.
The reference methods are nicMSlesion by \cite{valverde2019one}, DeepMedic by \cite{kamnitsas2017efficient}, and 2.5D Tiramisu by \cite{zhang2019multiple}. These methods have been selected for the availability of the authors source code and the relevance of their contributions in the MS segmentation community.  

\textbf{nicMSlesion:} This method is based on a cascade of two 3D patch-wise CNNs. The first one is trained to be sensitive in order to reveal lesion candidates. The second one is trained to reduce the misclassified voxels from the first network. 
Training is performed on $11\times11\times11$ patches randomly augmented with flipping and rotations. Therefore, nicMSlesion involves classical data augmentation. In the first network, the negative class is under-sampled to the same number of existing lesion voxels. It is composed of patches extracted from all of the available lesion voxels and a random selection of normal appearing tissue voxels. Afterwards, an evaluation of the first CNN model is computed by performing inferences on the same train-set and identifying negative voxels that have been misclassified as lesions (False Positives). Finally, the second model is trained using a balanced set composed of all the lesion voxels and a random selection from the identified False Positives in the previous step. 

\textbf{DeepMedic:}
This method is based on a 11-layers deep dual pathway 3D CNN designed for brain lesion segmentation. 
In order to incorporate both local and larger contextual information, the dual pathway architecture processes the input images at multiple scales simultaneously.
To overcome the computational burden, the authors use a hybrid dense training scheme  processing adjacent image patches into one pass through the network.
To refine the network segmentation and remove false positives, a 3D fully connected conditional random field is used. 
The training includes data augmentation with sagittal reflections.

\textbf{2.5D Tiramisu:} This method is based on a fully convolutional densely connected network. The model uses stacked slices from all three anatomical planes to achieve a 2.5D based method. Individual slices from a given orientation provide global context along the plane and the stack of adjacent slices adds local context. 
The training also includes flipping and rotations of the 2D patches for data augmentation. Therefore, 2.5D Tiramisu involves classical data augmentation. 
For each stack of 2D $128\times128$ slices composed of a center slice and its 2 adjacent slices, the model produces the segmentation of the center slice. Then, the inference results along the different orientations are combined via majority vote to output the final segmentation.
\\

For both these methods, we use the implementations provided publicly by the authors (see \footnote{https://github.com/sergivalverde/nicMSlesions} and \footnote{https://github.com/MedICL-VU/LesionSeg} ).

\subsubsection{Statistical Test}
To assert the advantage of a technique obtaining the highest average score, we conducted a Wilcoxon test over the lists of hybrid scores measured at image level (for the consistency of the segmentations section we took the lists of dice indices between the two segmentations). The significance of the test is established for a p-value below 0.05.  In the following tables, * indicates a significantly better average score when compared with the rest of the other approaches.


\section{Results}


\subsection{Ablation Study}\label{abla}

To demonstrate the impact of each proposed contribution on domain generalization, we measured separately their effects on different metrics. To show both the effect on accuracy improvement and the domain shift robustness, we propose out-of-domain and in-domain ablation study. First, we trained each method configuration on ISBI challenge train-set, then we validated on both ISBI test-set (see Table \ref{tab:ablation_same}) and MSSEG'16 (see Table \ref{ablation}).
To ensure a fair comparison, each configuration is trained until convergence. Specifically, we used an early stopping criterion of 50 epochs (\textit{i.e.}, the training stops if the loss function does not improve on the validation set during 50 epochs) with a maximum number of 500 epochs. We verified that none of the configurations reached this maximum number.

\begin{table*}[htb!]
\centering
\caption{Ablation study results with different variants of our approach trained on ISBI challenge train-set and tested on ISBI test-set. DeepLesionBrain (DLB) refers to using our spatially distributed specialized networks, each network in charge of segmenting a sub-volume. The generic network represents the variant of DLB with a single network (without the spatially distributed strategy). Hierarchical Specialized Learning (HSL) indicates that we initialized the “specialized networks” with the “generic Network”. To evaluate the performance of the proposed Data Augmentation, we compared variants with IQDA (previously defined in \ref{IQDA}) and without IQDA. For each metric, the bold values indicate the best result. In hybrid score column, * indicates a significantly better score than the other approaches using the Wilcoxon test.}

\label{tab:ablation_same}
\resizebox{450pt}{!}{%
\begin{tabular}{|c|c|c|c|c|c|c|c|}
\hline
                Method          &Hybrid Score    & Dice           & PPV            & TPR            & LFPR           & LTPR & Submission Score                       \\ \hline
DLB with HSL and IQDA   & \textbf{0.747*} & 0.646   & 0.888  & 0.545   & 0.131  & 0.486  & \textbf{92.849}  \\ 
DLB with HSL and without IQDA   & 0.732  & \textbf{0.677}  & 0.849 & 0.603  & 0.192  & \textbf{0.489} & 92.383  \\
DLB  without HSL and with IQDA &0.710  & 0.576    & \textbf{0.892} & 0.453  & \textbf{0.121} & 0.360  & 91.713  \\ 
DLB  with models genesis init. and IQDA & 0.718  & 0.621  & 0.867 & 0.513  & 0.187 & 0.438  & 91.885  \\
DLB  with AssemblyNet init. and IQDA & 0.723  & 0.628
    & 0.885 & 0.515 & 0.140 & 0.406  & 92.109  \\
The generic network with IQDA    & 0.736 & 0.668   & 0.859   & 0.585  & 0.178  & \textbf{0.489}  & 92.491 \\ 
The generic network without IQDA    & 0.688 & 0.654 & 0.502    & \textbf{0.869} & 0.162     & 0.468 & 92.425 \\ 
 
\hline
\end{tabular}%
}
\end{table*}

Table \ref{tab:ablation_same} shows the effect of each contribution to segmentation accuracy, when trained on ISBI challenge train-set and tested on ISBI test-set.
First, the best performing combination is DLB with HSL and IQDA, it obtained an ISBI Score of 92.849. 
Second, both the versions of DLB without IQDA and DLB without HSL are less accurate. They obtained respectively ISBI scores of 92.383 and 91.713. The later comparison shows the impact of HSL on the accuracy of segmentations.
Moreover, the generic network is less accurate than our spatially distributed approach used in DLB. The variant of generic Network with IQDA obtained a score of 92.491, whereas the variant without IQDA obtained a hybrid score of 92.425.
Finally, we compare HSL with other weight initialization strategies. Specifically, HSL is compared with the neighbor transfer learning from AssemblyNet proposed by \cite{coupe2020assemblynet} and models genesis proposed by \cite{zhou2021models}.
Although both variants obtained a better score compared to DLB without HSL, both initialization strategies gave a lower score than DLB with HSL and IQDA. 

\begin{table*}[htb!]
\centering
\caption{Ablation study results with different variants of our approach trained on ISBI challenge train-set and tested on MSSEG'16 (see caption of Table \ref{tab:ablation_same} for details).}
\label{ablation}
\resizebox{380pt}{!}{%
\begin{tabular}{|c|c|c|c|c|c|c|}
\hline
Method   & Hybrid Score   & Dice   & PPV & TPR & LFPR    & LTPR  \\\hline
DLB with HSL and IQDA  & \textbf{0.684*} & 0.639 & 0.768& 0.608 & \textbf{0.319}     & 0.700  \\
DLB with HSL and without IQDA   & 0.673   & \textbf{0.669}    & 0.728   & 0.671   & 0.416   & 0.725        \\
DLB without HSL and with IQDA & 0.648   & 0.562   & \textbf{0.806} & 0.489   & 0.320 & 0.629   \\
DLB with models genesis init. and IQDA &0.623 &0.593& 0.737& 0.576& 0.436& 0.665     \\
DLB with AssemblyNet init. and IQDA &0.610& 0.541& 0.708& 0.537& 0.466  & 0.705   \\
The generic Network with IQDA   & 0.672  & 0.665    & 0.721    & \textbf{0.673}   & 0.413  & \textbf{0.727}    \\
The generic network without IQDA  & 0.626  & 0.625 & 0.763 &  0.588 & 0.449  & 0.611    \\

 \hline
\end{tabular}%
}
\end{table*}

Table \ref{ablation} shows the effect of each contribution to domain shift robustness, when trained on ISBI challenge train-set and tested on MSSEG'16.
First, the most robust combination is DLB with HSL and IQDA, it obtained a hybrid score of 0.684. 
Second, both the variants of DLB without IQDA and DLB without HSL are less accurate. They obtained hybrid scores of 0.673 and 0.648 respectively.
Moreover, the generic network is less robust than our spatially distributed approach with DLB. The variant of the generic network with IQDA obtained a score of 0.672, whereas the variant without IQDA obtained a hybrid score of only 0.626. The later comparison shows the impact of IQDA on robustness even without the spatially distributed networks. 
Finally, HSL is compared with Assembynet \cite{coupe2020assemblynet} and model genesis \cite{zhou2021models} initialization strategies. The variants with model genesis and AssemblyNet initialization methods obtained respectively hybrid scores of 0.623 and 0.6103.

\subsection{Cross-dataset Testing}\label{Cross Dataset Testing}
In this section, we assess the cross-dataset robustness and generalization ability of our proposed approach. We chose to compare our method with two state-of-the-art approaches:
nicMSlesion \cite{valverde2019one}, DeepMedic \cite{kamnitsas2017efficient}, and 2.5D Tiramisu \cite{zhang2019multiple}.\\
During the proposed validation, all the methods have been trained on exactly the same dataset (\textit{i.e.}, same preprocessing, same number of modalities, \textit{etc.}) to ensure a fair comparison of method performance. Although refernece methods have been originally proposed with a specific number of modalities (\textit{i.e.}, Tiramisu 2.5D and nicMSlesion were tested with 4 and 3 modalities respectively), their implementation is independent of the number of modalities since all modalities are concatenated and feeded to the CNN.

Besides, their official open source implementations support the usage of only T1w and FLAIR sequences. Thus, in this evaluation all methods are trained using only these two modalities. 
The following cross-dataset testing (cross-domain testing) consists in training each technique on one dataset at each time. Afterwards, the obtained models are evaluated on the other datasets which contain unseen domains.
We verified the average inference time per image for each method on the same machine: 57.353s for nicMSlesion, 17.547s for DeepMedic, 47.471s for 2.5D Tiramisu, and 38.014s for DLB (this time does not include preprocessing since similar for all the methods). 
Unlike using a single network to segment patches coming from the entire image, DLB uses multiple networks, one network for each brain area. These networks are loaded one by one to enable the use of a common GPU hardware solution (\textit{e.g.}, NVIDIA Titan Xp with 12 GB in our setup). Even though DLB requires sequential loading of multiple networks on GPU, the inference time over the whole image is similar to using a single network since network weights loading time is negligible compared to patch segmentation time.

The ISBI score is returned by the challenge website only for ISBI test-set evaluation, and thus this metric is not available (NA) for testing on other datasets.

\subsubsection{Trained on ISBI}

\begin{table*}[htb!]
\centering
\caption{Results of the different approaches trained on the ISBI training dataset, with T1w and FLAIR modalities. For each metric, the bold values indicate the best result. In hybrid score column, * indicates a significantly better score than the three other approaches using the Wilcoxon test. Red values indicate hybrid scores lower than 0.5 or Dice index below 0.25.}
\label{tab:isbi}
\resizebox{\textwidth}{!}{%
\begin{tabular}{|c|c|c|c|c|c|c|c|c|c|}
\hline
Trained on ISBI                    & Approach    & Hybrid Score    & Dice           & PPV            & TPR            & LFPR           & LTPR           & CORR & Submission Score               \\ \hline

\multirow{4}{*}{MSSEG'16}        & nicMSlesion & 0.537           & 0.442          & 0.614          & 0.423          & 0.504          & 0.629          & 0.495  & NA                     \\
& DeepMedic & 0.510& 0.476 & 0.542 & 0.560 & 0.829 & \textbf{0.850}  & 0.509                     & NA   \\
                                   & 2.5D Tiramisu    & \textbf{0.711}  & \textbf{0.664} & 0.741          & \textbf{0.658} & \textbf{0.284} & 0.695          & \textbf{0.730}  & NA             \\
                                   & DLB         & 0.684           & 0.639          & \textbf{0.768} & 0.608          & 0.319          & 0.700 & 0.650 & NA                       \\ \hline
\multirow{4}{*}{In-house dataset}  
& nicMSlesion & {\color[HTML]{FF0000} 0.419} & {\color[HTML]{FF0000} 0.204} & 0.727 & 0.129 & 0.309 & 0.361  & 0.158                     & NA   \\

& DeepMedic & 0.523& 0.536 & 0.633 & 0.499 & 0.805 & \textbf{0.765}  & 0.549                     & NA   \\

& 2.5D Tiramisu    & 0.654           & 0.545          & \textbf{0.871} & 0.410          & \textbf{0.204} & 0.476          & 0.635                     & NA   \\
                                   & DLB         & \textbf{0.696*} & \textbf{0.675} & 0.850          & \textbf{0.564} & 0.342          & 0.644 & \textbf{0.718}   & NA          \\ \hline
\end{tabular}%
}
\end{table*}

Table \ref{tab:isbi} shows the results of segmentation  when training the different approaches using T1w and FLAIR modalities, on the ISBI training dataset (2D resolution FLAIR). 
\\
When validating the methods on MSSEG'16, we report that 2.5D Tiramisu obtained slightly better results (not significantly) than DLB, in terms of hybrid score whereas nicMSlesion and DeepMedic performed relatively worse with 0.537 and 0.51 respectively. \\
On our in-house dataset, DLB performed significantly better with a hybrid score of 0.696 while 2.5D Tiramisu, DeepMedic, and nicMSlesion obtained respectively 0.654, 0.523, and 0.419.
We can notice that nicMSlesion offers poor cross domain performance on 3D FLAIR when trained with 2D FLAIR dataset.

\subsubsection{Trained on MSSEG'16}
\begin{table*}[htb!]
\centering
\caption{Results of the different approaches trained on the MSSEG'16 dataset, with T1w and FLAIR modalities. For each metric, the bold values indicate the best result. In hybrid score column, * indicates a significantly better score than the three other approaches using the Wilcoxon test. 
Red values indicate hybrid scores lower than 0.5 or Dice index below 0.25. }
\label{tab:msseg}
\resizebox{\textwidth}{!}{%
\begin{tabular}{|c|c|c|c|c|c|c|c|c|c|}
\hline
Trained on MSSEG'16                  & Approach    & Hybrid Score    & Dice           & PPV            & TPR            & LFPR           & LTPR           & CORR           & Submission Score      \\ \hline
\multirow{4}{*}{ISBI test-set} & 
nicMSlesion & 0.555           & 0.398          & 0.717          & 0.292          & 0.368          & 0.206          & 0.822          & 87,173          \\

& DeepMedic & 0.547 & 0.378 &0.801&	0.265&	0.416&	0.298& 0.717          & 87.344   \\

                                   & 2.5D Tiramisu    & {\color[HTML]{FF0000} 0.462}           & {\color[HTML]{FF0000} 0.165}          & \textbf{0.937} & 0.096          & \textbf{0.075} & 0.160          & 0.212          & 86,686          \\
                                   & DLB         & \textbf{0.618*} & \textbf{0.535} & 0.697          & \textbf{0.471} & 0.353          & \textbf{0.373} & \textbf{0.835} & \textbf{89.043} \\ \hline
\multirow{4}{*}{In-house dataset}  & nicMSlesion & 0.669           & 0.686          & 0.689          & 0.705          & 0.467          & 0.717          & 0.737          & NA              \\
& DeepMedic & 0.597 & 0.645& 0.647& 0.670& 0.721& \textbf{0.811}& 0.650          & NA   \\
                                   & 2.5D Tiramisu    & 0.664           & 0.706          & \textbf{0.766} & 0.694          & \textbf{0.432} & 0.801 & 0.552          & NA              \\
                                   & DLB         & \textbf{0.697*} & \textbf{0.746} & 0.681          & \textbf{0.847} & 0.478          & 0.754          & \textbf{0.799} & NA              \\ \hline
\end{tabular}%
}
\end{table*}

Table \ref{tab:msseg} shows the results of segmentation when training the different approaches on the MSSEG'16 dataset comprising 3D T1w and 3D FLAIR modalities.
First, we notice that our approach obtained significantly better hybrid scores for both ISBI test and In-house  datasets. \\
Second, when validating on ISBI, the obtained submission score is 89.043 for DLB (the closest to human performance), 87.344 for DeepMedic, 87.173 for nicMSlesion, and 86.686 for 2.5D Tiramisu (the farest from human performance). In the same conditions, 2.5D Tiramisu obtained the average Dice of 0.165 that indicates a failure of the method and thus a lack of generalization in this scenario (when trained on high quality 3D FLAIR and tested on low quality 2D FLAIR).\\
Finally for In-house dataset, DLB produced significantly better segmentation than other methods. DLB obtained a hybrid score of 0.697 while nicMSlesion obtained 0.669, 2.5D Tiramisu obtained 0.664, and DeepMedic obtained the lowest score of 0.597. 

\subsubsection{Trained on In-house}
\begin{table*}[htb!]
\centering
\caption{Results of the different approaches trained on In-house dataset, with T1w and FLAIR modalities. For each metric, the bold values indicate the best result. In hybrid score column, * indicates a significantly better score than the three other approaches, using the Wilcoxon test. Red values indicate hybrid scores lower than 0.5 or Dice index below 0.25.}
\label{tab:In-house}
\resizebox{\textwidth}{!}{%
\begin{tabular}{|c|c|c|c|c|c|c|c|c|c|}
\hline
Trained on In-house    & Approach    & Hybrid Score    & Dice           & PPV            & TPR            & LFPR           & LTPR           & CORR           & Submission Score      \\ \hline
\multirow{4}{*}{MSSEG'16}    & nicMSlesion & 0.700           & 0.650          & \textbf{0.822} & 0.586          & \textbf{0.150} & 0.607          & 0.607          & NA              \\

& DeepMedic & 0.717& 0.694 & 0.750& 0.701& 0.345& \textbf{0.782} & 0.709   & NA   \\
                     
 & 2.5D Tiramisu    & \textbf{0.745}  & 0.665          & 0.741          & 0.687          & 0.164          & 0.720 & 0.722          & NA              \\
                               & DLB         & 0.741           & \textbf{0.719} & 0.735          & \textbf{0.744} & 0.209          & 0.671          & \textbf{0.776} & NA              \\ \hline
\multirow{4}{*}{ISBI test-set} & nicMSlesion & {\color[HTML]{FF0000} 0.453}           & {\color[HTML]{FF0000} 0.131}          & 0.644          & 0.075          & 0.338          & 0.050          & 0.712          & 84,512          \\
                               
& DeepMedic &0.523 & 0.385&		0.807&	0.273&	0.388&	\textbf{0.215} &0.670
   & 86.810  \\

                               & 2.5D Tiramisu    & 0.608           & 0.355          & \textbf{0.938} & 0.231          & \textbf{0.065} & 0.160          & 0.689          & 89.289          \\
                               & DLB         & \textbf{0.638*} & \textbf{0.476} & 0.877          & \textbf{0.348} & 0.104          & 0.193 & \textbf{0.787} & \textbf{89.843} \\ \hline

\end{tabular}%
}
\end{table*}

Table \ref{tab:In-house} shows the results of segmentation when training on our In-house dataset with 3D FLAIR. 
First, the obtained results when testing on MSSEG'16 indicates a close segmentation accuracy for DLB and 2.5D Tiramisu in terms of hybrid score (0.745 and 0.741), and a slightly lower performance from nicMSlesion and DeepMedic (0.7 and 0.717). 
Second, we notice that our approach obtained a significantly higher hybrid score when validating on the ISBI testing dataset, with a submission score of 89.843 compared to 2.5 Tiramisu, DeepMedic, and nicMSlesion with 89.289, 86.810, and 84.512 respectively. In this scenario, nicMSlesion obtained the worst score with a dice of 0.131 indicating a failure of the method.

\subsubsection{Cross-dataset Testing Summary}

First, it is noteworthy that when our approach obtained a better score, the superiority was statistically significant. On the contrary, when one of the other approaches obtained a higher score, the advantage was not significant using the Wilcoxon test.
Second, it should be pointed out that in all the considered cross-domain cases, DLB did not degenerate not even once while maintaining high scores. We reported for nicMSlesion trained on ISBI and validated on the In-house dataset a hybrid score of 0.419. We also recall the low performance of 2.5D Tiramisu trained on MSSEG'16 and tested on ISBI (0.462 hybrid score). This shows the cross-domain robustness of the proposed strategy.

Table \ref{tab:Cross-Dataset Summary} sums up cross-dataset experiments results. This table presents the average score estimated over all the images obtained during the three experiments presented in Table \ref{tab:isbi}, Table \ref{tab:msseg}, and Table \ref{tab:In-house} (61 images for ISBI test-set, 43 images for In-house, 15 images for MSSEG'16). We notice that DLB obtains the highest hybrid score and Dice index by a large margin compared to 2.5D tiramisu, DeepMedic, and nicMSlesion.

\begin{table*}[htb!]
\centering
\caption{Summary of the cross-dataset experiment. The table represent the average of cross-dataset experiment results (see Table \ref{tab:isbi}, Table \ref{tab:msseg}, and Table \ref{tab:In-house}) based on the number of images for each dataset. For each metric, the bold values indicate the best result. In hybrid score column,  * indicates a significantly better score than the three other approaches using the Wilcoxon test.}
\label{tab:Cross-Dataset Summary}
\resizebox{400pt}{!}{%
\begin{tabular}{|c|c|c|c|c|c|c|c|}
\hline
Strategy       & Hybrid Score    & Dice           & PPV            & TPR            & LFPR           & LTPR           & CORR           \\ \hline

nicMSlesion    & 0.526           & 0.365          & 0.695          & 0.308          & 0.362          & 0.338          & 0.595          \\
DeepMedic &  0.554&	0.483&	0.725&	0.429&	0.556&	\textbf{0.520}&	0.649\\
2.5 D Tiramisu & 0.608           & 0.443          & \textbf{0.870} & 0.368          & \textbf{0.179} & 0.402          & 0.537          \\
DLB            & \textbf{0.663*} & \textbf{0.601} & 0.775          & \textbf{0.550} & 0.299         & 0.484 & \textbf{0.780} \\ \hline

\end{tabular}%
}
\end{table*}


\subsection{Same Domain Validation} \label{same domain val}
Despite the previously mentioned limitations of in-domain validation, we also provide experiments using the same domain as complementary results.
First, Table \ref{tab:isbi_same} shows the results of DLB, nicMSlesion, DeepMedic, and 2.5D Tiramisu on ISBI test-set after being trained on ISBI train-set (same domain), with T1w and FLAIR modalities. The three approaches give close results with submission scores of 92.923 for 2.5D Tiramisu, 92.849 for DLB, and 92.161 for nicMSlesion. DeepMedic comes last with a submission score of 90.866.

Second, Table~\ref{tab:soa} shows the current top performing methods on the ISBI challenge website. 2.5D Tiramisu \cite{zhang2019multiple} is the best ranked method with the current highest ISBI Score of 93.21, followed in second place by nnUnet \cite{isensee2019nnu} with 93.09. Both approaches rely on 4 modalities (T1w, FLAIR, T2, PD). Our approach comes in third place using only 2 modalities, and obtained the ISBI submission score of 92.85.
Although DLB uses a lower number of modalities, it obtained better results than IMAGINE \cite{hashemi2018asymmetric}, Self-adaptive network \cite{feng2019self}, and Multi-branch \cite{aslani2019multi} that obtained respectively the scores of 92.49, 92.41, and 92.12.\\

\begin{table*}[ht!]
\centering
\caption{Results of the different approaches trained on the ISBI training dataset and tested on ISBI test-set, with T1w and FLAIR modalities. For each metric, the bold values indicate the best result. In hybrid score column, * indicates a significantly better score than the three other approaches using the Wilcoxon test.}
\label{tab:isbi_same}
\resizebox{480pt}{!}{%
\begin{tabular}{|c|c|c|c|c|c|c|c|c|}
\hline
 Approach    & Hybrid Score    & Dice           & PPV            & TPR            & LFPR           & LTPR           & CORR           & Submission Score      \\ \hline
  nicMSlesion & 0.724           & 0.639          & 0.853          & 0.541          & 0.144          & 0.432          & 0.863          & 92.161          \\
  
 DeepMedic& 0.649&	0.643&	0.827&	0.557&	0.408&	\textbf{0.530}&	\textbf{0.873}&	90.866 \\
                                    2.5D Tiramisu    & \textbf{0.750}  & \textbf{0.672} & 0.865          & \textbf{0.592} & 0.150          & 0.513 & 0.868 & \textbf{92.923} \\
                                  DLB         & 0.748           & 0.646          & \textbf{0.888} & 0.545          & \textbf{0.131} & 0.486          & 0.868 & 92.849          \\ \hline

\end{tabular}%
}
\end{table*}
\begin{table*}[ht!]
\centering
\caption{State-of-the-art published results for the ISBI challenge.
}
\label{tab:soa}
\resizebox{440pt}{!}{%
\begin{tabular}{|c|c|c|c|}
\hline

Approach & Modalities  & CNN type & Submission Score             \\ \hline

2.5D Tiramisu \cite{zhang2019multiple} & T1w, FLAIR, T2, PD     & 2D   & \textbf{93.21}       \\ \hline
nnUnet \cite{isensee2019nnu} & T1w, FLAIR, T2, PD& 2D and 3D  & 93.09       \\ \hline
DLB [ours] & T1w, FLAIR  & 3D        & 92.85                  \\ \hline
IMAGINE \cite{hashemi2018asymmetric}& T1w, FLAIR, T2, PD  & 3D    & 92.49  \\ \hline
Self-adaptive network \cite{feng2019self}& T1w, FLAIR, T2, PD & 3D  & 92.41   \\ \hline
Multi-branch \cite{aslani2019multi}& T1w, FLAIR, T2       & 2D         & \textit{92.12}  \\ \hline
\end{tabular}%
}
\end{table*}

The high accuracy of the results were expected as both the training and testing sets share the same domain (same acquisition conditions, and same scanner...). By tuning and adapting a method to this specific domain conditions, we can obtain artificially higher performance (\textit{e.g.}, DLB with 4 modalities obtained a score of 92.92, and a 2D version of DLB obtained 93.14 \footnote{https://smart-stats-tools.org/lesion-challenge}).
However, in our opinion, results reported in the same domain experiment do not truly reflect methods performances. For instance, the best performing method of this section (2.5D Tiramisu) failed when trained on different datasets (obtained submission scores of 89.043 and 89.289 in Table \ref{tab:msseg} and Table \ref{tab:In-house}).
The limitation of such a validation strategy is one of the main messages of our paper. Hence, we consider that cross-dataset evaluation with diverse images from different domains is a better alternative for method assessment.

\subsection{Corss-dataset Segmentation Consistency}
Finally, a usually under-investiguated method property is its cross-dataset segmentation consistency. In order to assess the consistency of our model segmentation, we decided to compare the segmentation produced by each approach on the same data when the model is trained on different datasets. We compute the Dice between the different segmentations of a method as a similarity index to quantify the prediction consistency. Table \ref{tab:cross-dataset consistency} shows the segmentation consistency for each approach in our cross-dataset setting.

First, we analyzed the segmentations on In-house when the models are trained respectively on ISBI train-set and and MSSEG'16. In this case, DLB obtained the best score of 0.647, followed by 2.5D Tiramisu and DeepMedic with 0.6261 and 0.602 respectively. Lastly, nicMSlesion obtained a score of 0.217. Second, we analyzed the segmentations on MSSEG'16 when the models are trained respectively on ISBI train-set and In-house. In this case, we obtained close consistency scores for 2.5D Tiramisu and DLB with Dice scores around 0.72 while DeepMedic and nicMSlesion are less consistent with 0.537 and 0.514 respectively.
Finally, we analyzed the segmentations on ISBI test-set when comparing the models trained on ISBI train-set, the models trained on In-house, and the models trained on MSSEG'16. For all settings, DLB was significantly more consistent than both other methods with a Dice ranging from 0.63 to 0.649. 2.5D Tiramisu segmentation consistency index varies from 0.217 to 0.485. DeepMedic consistency index fluctuates from 0.49 to 0.602. nicMSlesion is the least consistent with scores ranging from 0.177 to 0.512. During our cross-dataset consistency experiment, DLB was the only method capable of ensuring segmentation consistency independent of the training dataset. Both other methods failed several times as indicated with red color in Table \ref{tab:cross-dataset consistency}.

\begin{table*}[htb!]
\centering
\caption{The consistency of the segmentations for each approach in the cross-dataset setting. The consistency index represents the test-set average of Dice values, each Dice is computed between two segmentations produced by the same method when trained on two different train-sets. Higher values indicate better consistency in the segmentations. The bold values indicate the best result and red values indicate consistency lower than 0.5. * indicates a significantly better segmentation consistency score than the three other approaches, using the Wilcoxon test.}
\label{tab:cross-dataset consistency}
\resizebox{\textwidth}{!}{%
\begin{tabular}{|cc|c|c|c|c|c|}
\hline
\multicolumn{2}{|c|}{Test-set}                                                                                                                                                                                  & In-house                           & MSSEG'16                          & \multicolumn{3}{c|}{ISBI Test-set}                                                                      \\ \hline
                                                                                                                                               &                                                                                    & \multicolumn{1}{l|}{ISBI Train-set} & \multicolumn{1}{l|}{ISBI Train-set} & \multicolumn{1}{l|}{ISBI Train-set} & In-house                     & \multicolumn{1}{l|}{ISBI Train-set} \\
\multirow{-2}{*}{Train-sets}                                                                                                                     & \multirow{-2}{*}{\begin{tabular}[c]{@{}c@{}}Dataset 1\\ vs. Dataset 2\end{tabular}} & MSSEG'16                          & In-house                           & MSSEG'16                          & MSSEG'16                    & In-house                           \\ \hline

 & nicMSlesion   & {\color[HTML]{FE0000} 0.217}  & 0.514  & 0.512  & {\color[HTML]{FE0000} 0.250} & {\color[HTML]{FE0000} 0.177}  \\
 
 & DeepMedic & 0.602& 0.537 & {\color[HTML]{FE0000}0.490} &0.602& {\color[HTML]{FE0000}0.496} \\
 
 & 2.5D Tiramisu  & 0.615   & \textbf{0.726} & {\color[HTML]{FE0000} 0.217}  & {\color[HTML]{FE0000} 0.485} &{\color[HTML]{FE0000} 0.460}       \\
\multirow{-4}{*}{\begin{tabular}[c]{@{}c@{}}The consistency of \\ the model predictions \\when trained on \\ Dataset1 vs. Dataset2\end{tabular}} 
 & DLB      & \textbf{0.647}     & 0.719         & \textbf{0.630*}      & \textbf{0.637*}   & \textbf{0.649*}   \\
\hline
\end{tabular}%
}
\end{table*}

\begin{figure*}[hbt]
  \centering
  \makebox[\textwidth]{\includegraphics[width=500px]{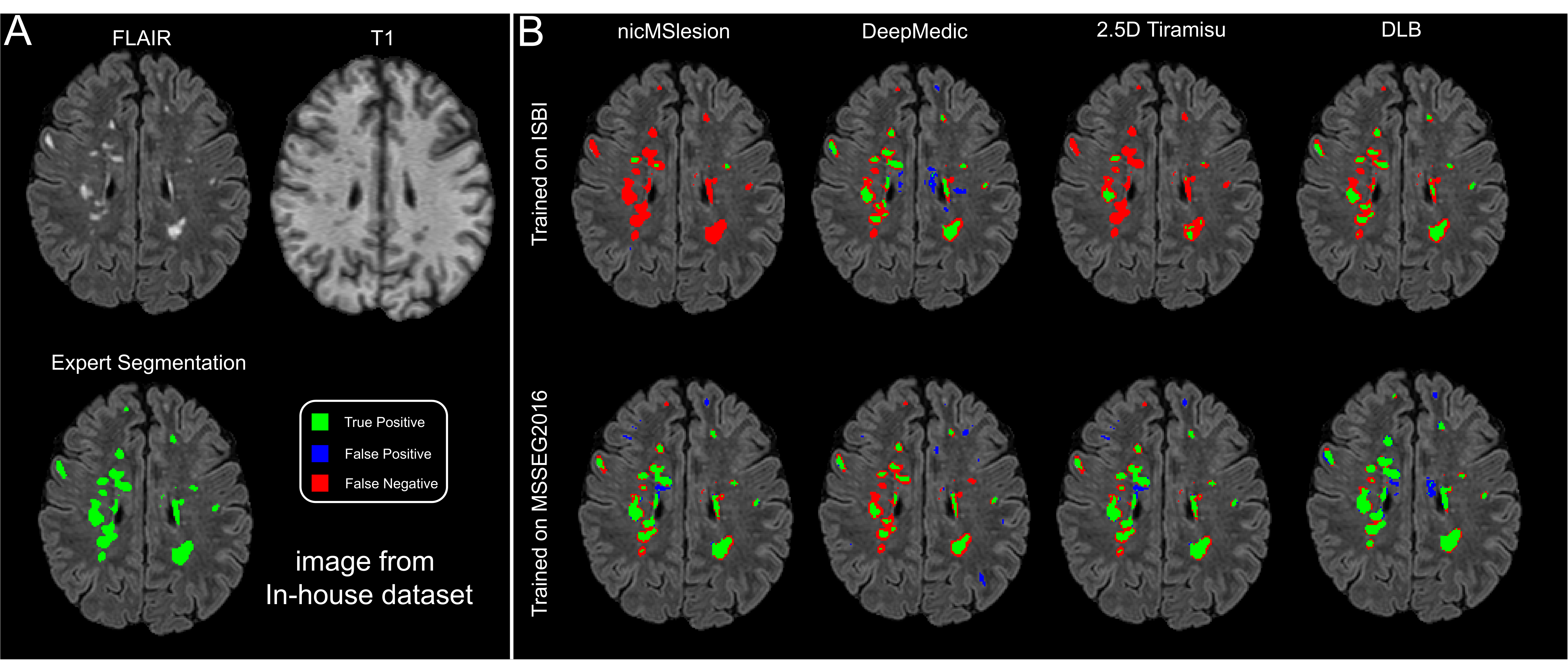}}
  \caption{Part A (left) axial sections of multi-modal MRI  (T1w and FLAIR) from In-house dataset, and its respective expert consensus segmentations for MS lesion segmentation. Part B (right) cross dataset segmentation of the image section shown in Part A. The first and second rows illustrate the segmentations of methods when trained respectively on ISBI dataset, and MSSEG'16 datasets. First, second, third, and fourth columns represent respectively the segmentations of nicMSlesion, DeepMedic, 2.5D Tiramisu, and DeepLesionBrain. }
  \label{fig:inhouse_sample}
\end{figure*}

\begin{figure*}[t!]
  \centering
  \makebox[ \textwidth]{\includegraphics[width=500px]{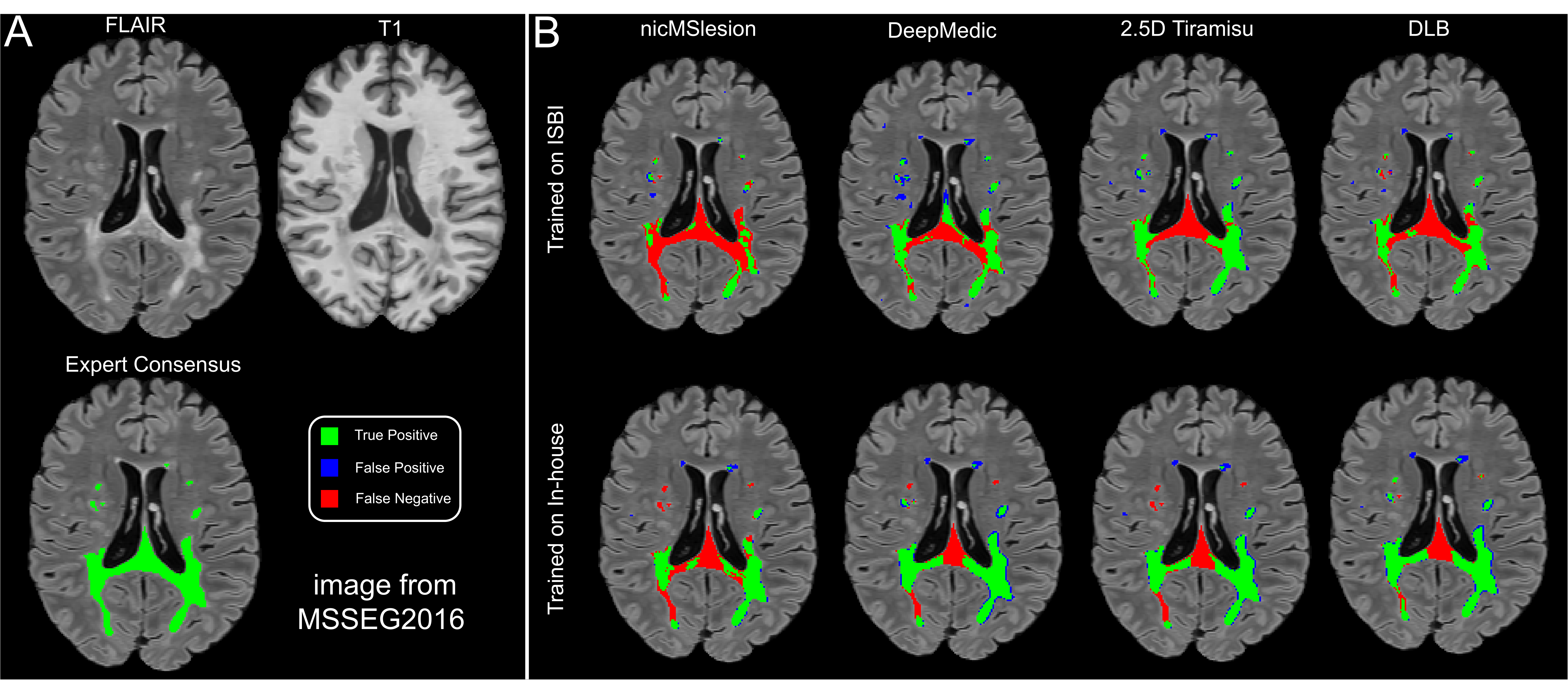}}
  \caption{Part A (left) axial sections of multi-modal MRI  (T1w and FLAIR) from MSSEG'16 dataset, and its respective expert consensus segmentations for MS lesion segmentation. Part B (right) cross dataset segmentation of the image section shown in Part A. The first and second rows illustrate the segmentations of methods when trained respectively on ISBI challenge, and In-house datasets. First, second, third, and fourth columns represent respectively the segmentations of nicMSlesion, DeepMedic, 2.5D Tiramisu, and DeepLesionBrain. }
  \label{fig:msseg_sample}
\end{figure*}

\begin{figure*}[t!]
  \centering
  \makebox[\textwidth]{\includegraphics[width=500px]{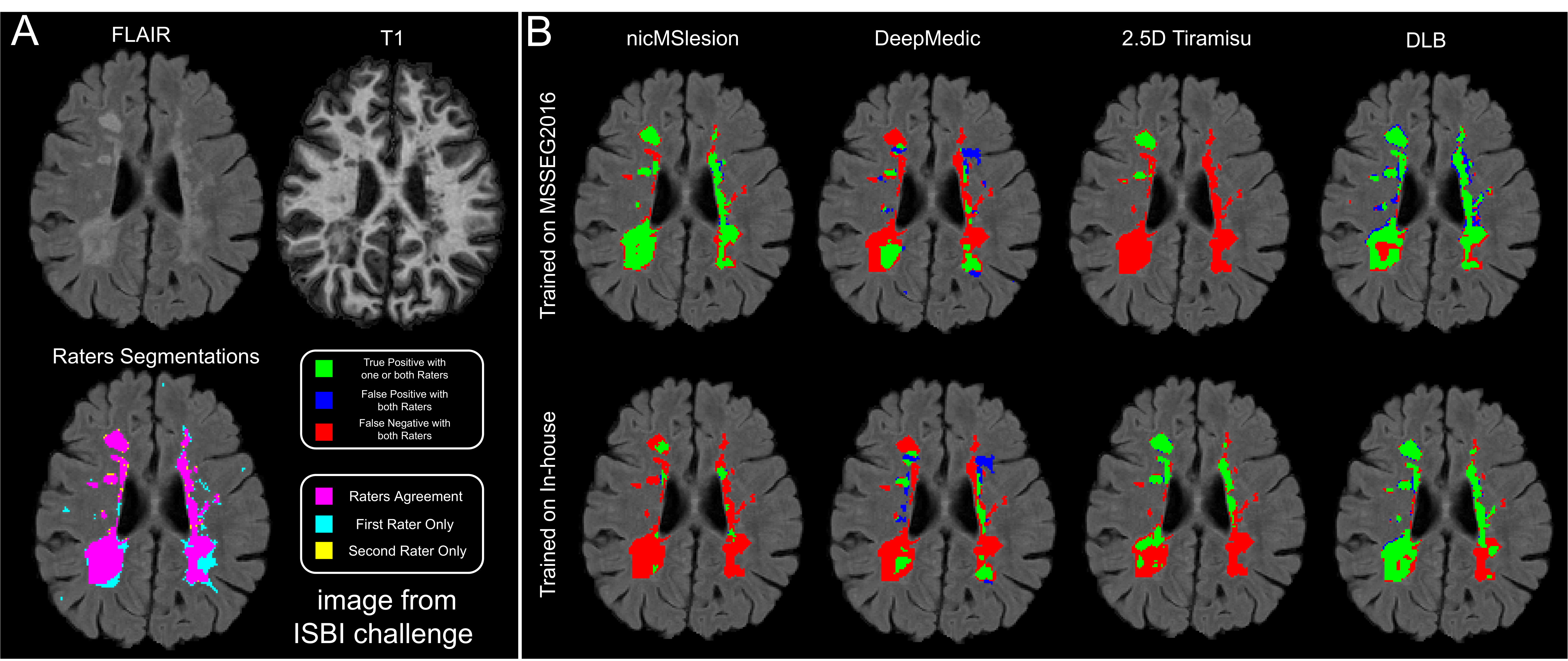}}
  \caption{Part A (left) axial sections of multi-modal MRI  (T1w and FLAIR) from ISBI challenge dataset, and its respective raters segmentations for MS lesion segmentation. Part B (right) cross dataset segmentation of the image section shown in Part A. The first and second rows illustrate the segmentations of methods when trained respectively on MSSEG'16, and In-house datasets. First, second, third, and fourth columns represent respectively the segmentations of nicMSlesion, DeepMedic, 2.5D Tiramisu, and DeepLesionBrain. }
  \label{fig:isbi_sample}
\end{figure*}
Figure \ref{fig:inhouse_sample} represents an image from the In-house dataset and the segmentation of the different methods when trained on ISBI challenge and MSSEG'16 datasets. First, both nicMSlesion and 2.5D Tiramisu fail to segment the majority of lesions when trained on ISBI challenge dataset. This exhibits the limitation of the robustness of these methods to domain shift, especially for 2.5D Tiramisu currently considered as the state-of-the-art approach on the ISBI challenge. Second, DLB detects almost all the lesions in the same conditions.
Third, although DeepMedic also detects most of the lesions, it is more prone to false positives compared to the other methods.
Finally, when choosing MSSEG'16 as a training dataset, DLB produces the most similar segmentation to expert annotation.  

Figure \ref{fig:msseg_sample} represents an image from MSSEG'16 dataset and the segmentation of the different methods when trained on ISBI challenge and In-house datasets. 
First, when trained on ISBI dataset, the segmentations of 2.5D Tiramisu, DeepMedic, and DLB are more accurate than nicMSlesion segmentation, although all techniques missed a large portion of the central lesion (False Negative) located around the midsagittal plane. 
These common voxels misclassification can result from the subjectivity of raters between training and testing datasets. 
Second, when trained on In-house, DLB delineates successfully most of the lesions. Especially in the case of small lesions, DLB misses only one lesion, whereas both nicMSlesion and 2.5 D Tiramissu miss four lesions, and DeepMedic misses two lesions.

\medskip
Figure \ref{fig:isbi_sample} represents an image from the ISBI challenge and the segmentation of the different methods when trained on MSSEG'16 and In-house datasets. From the four methods, DLB had the most consistent segmentation across different conditions of training domains. In this case, nicMSlesion produced a decent segmentation for this example only when trained on MSSEG'16. Likewise, 2.5D Tiramisu produced better segmentation when trained on In-house than on MSSEG'16. Although DeepMedic is consistent for this case, the produced segmentation is less precise and prone to false positives compared to the other methods.
\section{Discussion and Conclusion}

\subsection{Discussion}

Segmentation models trained with supervised learning can be sensitive to domain shift, that leads to the generalization failure. Such a domain shift may be caused by hardware and pre-processing diversity, difference in acquisition protocol or annotation protocol, that results in difference between the distributions of training and testing datasets. Besides, we also have to acknowledge the subjectivity of raters in training datasets. Indeed, the disagreement between expert segmentations, both in the same dataset and across different datasets, can make it difficult to train a generalizing model. Our experiments showed the limited generalization capability of state-of-the-art approaches, whereas DLB was able to adapt across different domains. Our study emphasizes the importance of cross-dataset validation, particularly when considering clinical application of machine learning.

DLB uses a group of several separately trained networks, each network is specialized in a particular region of the brain. This strategy makes our model less sensitive to domain shift, since the final segmentation is based on consensus of diversified and specialized networks. In the ablation study (see Tables \ref{tab:ablation_same} and  \ref{ablation}) , our spatially distributed networks strategy showed better generalization and higher accuracy than using a single model.

Automated MS lesion segmentation should be able to render the most accurate segmentation with the minimum number of modalities, in order to be efficiently adopted in clinical conditions and to limit inter-modality dependence. Many experts agree that FLAIR is the most important modality for MS lesion delineation. Moreover, T1w modality can provide complementary information for better white-matter, gray-matter, and cerebrospinal-fluid distinction. FLAIR and T1w are the most available modalities for MS patients and in all MS benchmark datasets. Our method achieved a competitive performance using these two modalities even on unseen domains.

In this paper, we proposed a novel data augmentation technique to reduce domain shift introduced by the variability of image resolution and quality. IQDA simulates different acquisition conditions in order to reduce covariate shift. Our ablation study (see Tables \ref{tab:ablation_same} and \ref{ablation}) showed IQDA as a solid contribution to segmentation accuracy and cross-domain generalization. 
Indeed, while other methods (nicMSlesion, DeepMedic, and 2.5D Tiramisu) involve usual data augmentation (rotation and flipping), such simple strategies failed to ensure good generalization on unseen dataset.

In our work we considered both specialized local features, and generic global features of MS lesions. 
The hierarchical specialization learning proposes an alternative to network cascades (\textit{i.e.}, \cite{valverde2017improving}). Instead of using cascades which are prone to error propagation, we suggested a logical hierarchy during learning based on data selection and transfer learning. The ablation study (see Tables \ref{tab:ablation_same} and  \ref{ablation}) exhibits the contribution of HSL to accuracy and domain generalization compared to DLB without HSL.

In this paper, the proposed method was validated using an out-of-domain cross-dataset evaluation. This strategy ensures that the performance obtained is not biased by the training dataset domain information. 
Indeed, the use of testing and training images from the same domain is questionable and does not reflect the generalization ability. The community should start considering this issue for both the validation and the comparison of proposed methods.

In section \ref{Cross Dataset Testing}, we reported that the best performances of DLB have been obtained when using high resolution 3D FLAIR datasets and multi-rater consensus ground truth for training. The resulting model is able to render more accurate segmentations for both 2D and 3D image resolution data, even across unseen domains. 
This observation led us to believe that to efficiently train 3D CNN based models for domain generalization, it may be desirable to optimize the model using high resolution training data.

With current available technology, it is unfeasible to exploit 3D CNNs with equivalent depth and kernel size as state-of-the-art 2D CNNs. Consequently, many neuroimaging automated pipelines are still using 2D CNNs despite processing 3D data. Our results suggest that using multiple compact networks can approximate a larger and more stable model since the sum of features extracted by the group of specialized networks and the features of a hypothetical big network may be equivalent in terms of relevant information for MS lesion segmentation. In our work, we have chosen to break down the complexity of the task spatially, based on the sub-volume division of the whole brain volume.
One other advantage of this distribution is the ability to train networks in parallel, since network weights and images of each region are independent. It is possible to use several GPUs for parallel training.

\subsection{Conclusion}
DeepLesionBrain is a deep learning framework for MS lesion segmentation designed for domain generalization. 
First, we use a spatially distributed strategy of multiple compact 3D CNNs with large overlapping receptive fields, in order to produce consensus based segmentation robust to domain shift. 
Second, we propose a novel image quality data augmentation to increase training data variability in a realistic way. 
Furthermore, we train our method using hierarchical specialization learning to efficiently incorporate both generic and specialized features. 
Finally, we use only T1w and FLAIR modalities to propose a method compatible with a large number of datasets.

The ablation study showed the impact of each contribution on segmentation accuracy and domain generalization. 
The out-of-domain cross-dataset testing is suggested as an alternative for method evaluation in areas which are sensitive to domain bias (\textit{i.e.}, medical imaging). Our validation showed the generalization ability of our method and its robustness to domain shift.
We also proved experimentally that DLB produces consistent segmentations compared to other state-of-the-art approaches regardless of the training data domain.

\section{Acknowledgements}
This work benefited from the support of the project DeepvolBrain of the French National Research Agency (ANR-18-CE45-0013). This study was achieved within the context of the Laboratory of Excellence TRAIL ANR-10-LABX-57 for the BigDataBrain project. Moreover, we thank the Investments for the future Program IdEx Bordeaux (ANR-10-IDEX-03-02, HL-MRI Project), Cluster of excellence CPU and the CNRS/INSERM for the DeepMultiBrain project. This study has also been supported by the DPI2017-87743-R grant from the Spanish Ministerio de Economia, Industria Competitividad. The authors gratefully acknowledge the support of NVIDIA Corporation with their donation of the TITAN Xp GPU used in this research.

\bibliographystyle{apalike}
\bibliography{paper}  


\end{document}